# Atomic structure generation from reconstructing structural fingerprints

Victor Fung,[1,4]* Shuyi Jia,[4] Jiaxin Zhang,[2] Sirui Bi,[2] Junqi Yin,[3] P. Ganesh,[1]

[1]*Center for Nanophase Materials Sciences, Oak Ridge National Laboratory, Oak Ridge, Tennessee 37831, United States*

[2]*Computer Science and Mathematics Division, Oak Ridge National Laboratory, Oak Ridge, Tennessee 37831, United States*

[3]*National Center for Computational Sciences, Oak Ridge National Laboratory, Oak Ridge, Tennessee*

[4]*School of Computational Science and Engineering, Georgia Institute of Technology, Atlanta, Georgia*

*E-mail: fungv@ornl.gov

## Abstract

Data-driven machine learning methods have the potential to dramatically accelerate the rate of materials design over conventional human-guided approaches. These methods would help identify or, in the case of generative models, even create novel crystal structures of materials with a set of specified functional properties to then be synthesized or isolated in the laboratory. For crystal structure generation, a key bottleneck lies in developing suitable atomic structure fingerprints or representations for the machine learning model, analogous to the graph-based or SMILES representations used in molecular generation. However, finding data-efficient representations that are invariant to translations, rotations, and permutations, while remaining invertible to the Cartesian atomic coordinates remains an ongoing challenge. Here, we propose an alternative approach to this problem by taking existing non-invertible representations with the desired invariances and developing an algorithm to reconstruct the atomic coordinates through gradient-based optimization using automatic differentiation. This can then be coupled to a generative machine learning model which generates new materials within the representation space, rather than in the data-inefficient Cartesian space. In this work, we implement this end-to-end structure generation approach using atom-centered symmetry functions as the representation and conditional variational autoencoders as the generative model. We are able to successfully generate novel and valid atomic structures of sub-nanometer Pt nanoparticles as a proof of concept. Furthermore, this method can be readily extended to any suitable structural representation, thereby providing a powerful, generalizable framework towards structure-based generation.

# Introduction

Finding materials with specified functional properties is a central goal in materials discovery and design, with important scientific and economic implications.[1-3] However, exhaustively evaluating materials within the accessible chemical space is often impractical, whether it is through experiments or computational simulations.[4] As an alternative to this high-throughput screening,[5] another tactic has been to design materials from the ground up with an explicit goal of meeting a target functionality.[6-9] This "inverse design" approach has the capability to greatly speeding up materials design by significantly reducing the candidates to test to only those which are likely to exhibit the desired properties.

In developing methods to generate new solid-state materials, it is not sufficient for the method to only provide the chemical composition of the material, but it should also specify its three-dimensional atomic structure as defined by their Cartesian coordinates.[10-11] The structure plays a key role in the property of a material, determining whether is it biologically viable as a drug, if it can catalyze a certain reaction, or if it can become superconducting, for example.[12-13] Traditionally, a host of algorithms have been developed for crystal structure prediction for a given composition by evaluating the atomic positions sampled through Bayesian optimization,[14] genetic algorithms,[15] swarm-based optimization,[16] and others.[17] However, efficiently finding global minimum solutions to the atomic structure remains difficult for high dimensional problems. Here, generative models[18-23] provide an attractive data-driven solution for inverse design due to their ability to learn the distribution of the chemical space that underlies a given functionality and then to efficiently sample from this distribution to predict new compounds.

While ML approaches can potentially offer a clear advantage, the problem of generating atomic structures with machine learning remains a challenging one, due to the Cartesian coordinates lacking the necessary rotational, translational, and permutational invariances for both efficient training and generation (Figure 1).[8, 24-25] Consequently, a number of approaches have been implemented to circumvent this limitation by sequentially building structures atom-by-atom,[26-27] applying an optimization process on pre-initialized positions through learned gradients,[28-30] or by generating an intermediate representation which are then reconstructed back to their positions.[20, 31-34]

We focus our discussion on the representation generation strategy as one of the most widespread approaches currently and allows for a direct one-shot sampling strategy without the need for iterative refinement. The most representative examples are those use in molecular generation, where the models generate representations in the form of strings such as SMILES[21] or as molecular graphs, which then encode back to the actual molecule. However, approaches using these representations do not explicitly encode the three-dimensional structure, and so have not been used beyond molecular systems. In going beyond molecular generation, choices remain limited as to which three-dimensional representations can be used, as they either satisfy the requisite invariances or are invertible, but rarely at the same time. Here, the most common examples use image or voxel-based representations where atoms are projected onto a three-dimensional grid.[20, 31, 33] Newly generated samples are then reconstructed from the voxel representation space back to their atomic positions. These approaches still suffer from disadvantages due to their lack of rotational invariances, lack of periodicity, high memory costs, and an unavoidable reconstruction loss due to the finite resolution of the grid.

Meanwhile, a large family of non-invertible structure representations satisfying these invariances have been developed over the years for use in machine learning potentials, such as the Behler-Parinello atom-centered symmetry functions (ACSF),[35] the smooth overlap of atomic positions (SOAP),[36] and others.[24, 37] While they can be extremely effective in capturing structural information for regression (i.e., the prediction of energy and forces), the lack of a closed-form solution to compute the Cartesian positions back from the representation had precluded its use in generative models up to this point. In this work we propose an iterative approach to solving for the positions which can be applied to any differentiable structure representation, thereby enabling them for use as the outputs of generative models. The central appeal in this approach lies in leveraging existing and highly efficient representations for structure generation, rather than attempting to design new representations which are often limited in effectiveness due to the condition of invertibility.

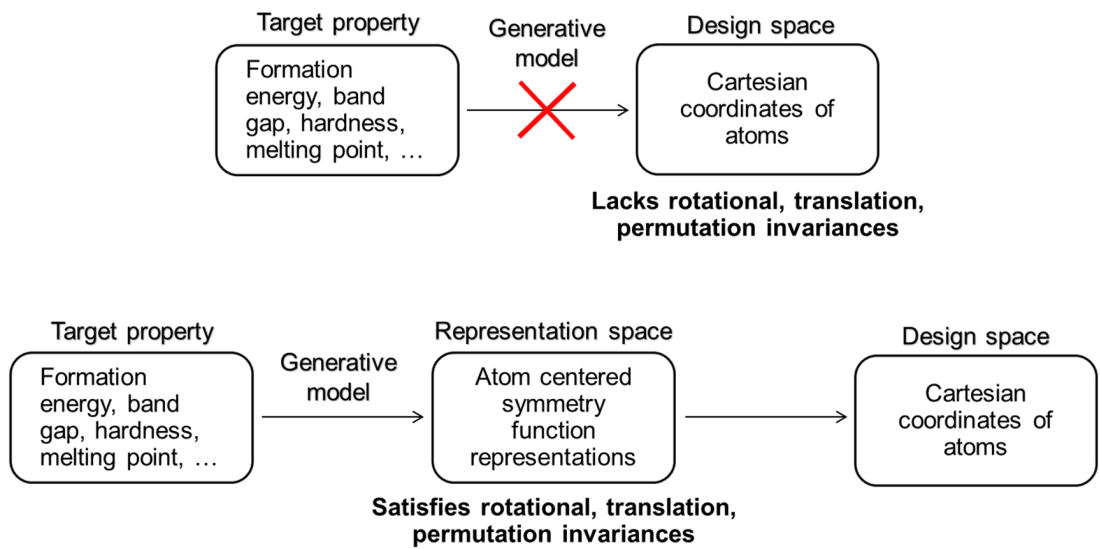

**Figure** 1: Scheme comparing direct generation of Cartesian coordinates versus a two-stage approach of generating within the representation space

## Methods

To reconstruct the atomic structure from a given representation, we treat it as a global optimization problem where the Cartesian positions of the atoms are the variables to be optimized and the objective is to minimize the difference or the loss function between the representation of the target structure and the ground truth solution (Figure 2). This is often a challenging problem where the global minimum cannot be easily found, especially for cases with many local minima. To significantly reduce the number of evaluations, we accelerate the optimization using gradients obtained from automatic differentiation, where the derivative of the loss function is taken with respect to the atomic positions. This assumes the representation is a function of the positions and is differentiable, which is almost universally true for these classes of structure representations. This yields a structure reconstruction workflow as follows. Given a representation R (to be reconstructed into Cartesian positions), the number of atoms N and the cell lattice parameters C, a separate set of atomic positions x are first randomly initialized with the same N and C. From the initialized positions, its representation R*(x) is computed, and the loss L between R and R* can be obtained. Here, both R*(x) and L(R*, R) are auto-differentiable functions, which allows us to obtain the derivative of L(R*, R) with respect to x. The positions x are then updated as x' with an optimizer using the gradients $s_x$ calculated via auto-differentiation. The procedure is repeated with R*(x) until a convergence criterion is met. The overall process is outlined in Algorithm 1.

**Algorithm 1** Structure reconstruction from its representation via gradient-based optimization
1: **Input:** representation R, the number of atoms N and (optionally) the dimensions of the periodic cell lattice parameters C
2. **while** initializations < max_ initializations **do**
3:   Randomly initialize a set of Cartesian positions $\{x_i\}_{i=1}^N$ within C
4.   **while** hops < max_hops **do**
5:     **while** iterations < max_iterations **do**
6:       calculate R*(x)
7:       calculate L(R*, R)
8:       obtain gradients $s_x$ from L(R*, R) using automatic differentiation
9:       use $s_x$ to update x to x' with optimizer'
10.     Add noise to x' to push positions out of local minima
11. Select x' with the lowest L(R*, R)
12: Construct crystal structure with x' and C

For a demonstration, we choose the classic ACSF[35] to compute the representation R. In this work, R is a concatenation of the outputs of the radial ACSF function $G^2$ and the angular ACSF function $G^5$:

$$G_i^2 = \sum_j e^{-\eta(D_{ij}-D_s)^2} \cdot f_c(D_{ij}) \tag{1}$$

$$G_i^5 = 2^{1-\zeta} \sum_{j,k \neq i}^{all} (1 + \lambda \cos \theta_{ijk})^\zeta e^{-\eta(D_{ij}-D_s)^2} \cdot f_c(D_{ij}) \cdot f_c(D_{ik}) \tag{2}$$

Here, $D_{ij}$ refer to the distances between atoms i and j, and $\theta_{ijk}$ refer to the angles between i, j, and k. A cutoff function $f_c$ is used where values become zero beyond the cutoff radius. The parameters, $\eta$, $D_s$, $\lambda$, and $\zeta$, for the symmetry functions used in this work are provided in the SI.

The ACSFs are able to provide local representations of each atomic environment, however a permutationally invariant global representation is desired for use in the generative model. To address this limitation, we then aggregate the representations over each atom in the system using max and min pooling. This method was chosen over simple averages to allow for greater variations in the values of R between structures, though other aggregation methods may also be used which are not exhaustively explored in this work. The final expression for R is then:

$$R = [max_{i=1}^N G_i^2, max_{i=1}^N G_i^5, min_{i=1}^N G_i^2, min_{i=1}^N G_i^5] \tag{3}$$

The ACSF functions were written in Pytorch[38] to calculate the gradients via auto-differentiation, which was based on an existing python implementation.[39] For the loss function

L(R*, R), we use the mean absolute error loss. We then use a gradient descent optimizer, Adam, to update x using $s_x$. The Adam optimizer,[40] which implements adaptive learning rates for each variable being optimized, has been highly effective in training neural network models and was found to be similarly effective for our problem.

It is important to note that with gradient descent alone, it is likely that converged solutions for x may fall into one of many local minima rather than the global optimum solution. Therefore, we incorporate multiple initializations with several "hops" within each initialization to more effectively sample through the local minima. Finally, the solution with the lowest loss is chosen, and the reconstructed atomic structure is then defined using the optimized x, and the constant C.

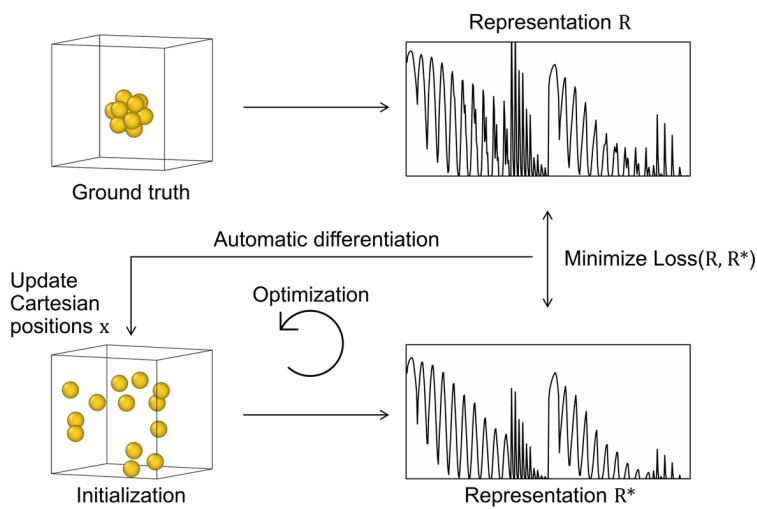

**Figure** 2: Scheme of the optimization algorithm for reconstructing Cartesian positions x from representation R

For the generative model, we use a conditional variational autoencoder (cVAE), which learns a Gaussian distributed latent representation z conditioned on y, following the approach by Sohn et al.[41] A loss function containing a reconstruction loss and a Kullback-Leibler loss is minimized in the training. Note this is not the same reconstruction loss referenced elsewhere in the work, which is associated with mapping the positions back from the representation. An additional parameter, beta, is used to adjust the weight of the two loss values,[42] which was set to be 0.75 with respect to the Kullback-Leibler loss. Here we use a cVAE with three hidden layers, a hidden dimension of 128, and a latent dimension of 25. The model was trained with the Adam optimizer with an initial learning rate of 0.001 for 2000 epochs and with a batch size of 64.

To quickly evaluate the generated samples for their formation energy, a surrogate model is trained which takes in R as the input, avoiding the need to first perform reconstruction to obtain x. The model is a feed-forward neural network with five layers and a hidden dimension of 256. The model was trained with the Adam optimizer with an initial learning rate of 5E-4 for 2000 epochs and with a batch size of 256. The training dataset was comprised of 100,000 structures described below, and the model obtains a MAE of 0.0028 eV/atom on the test split.

We find this accuracy to be sufficiently high to provide initial estimates of the quality of the generated samples.

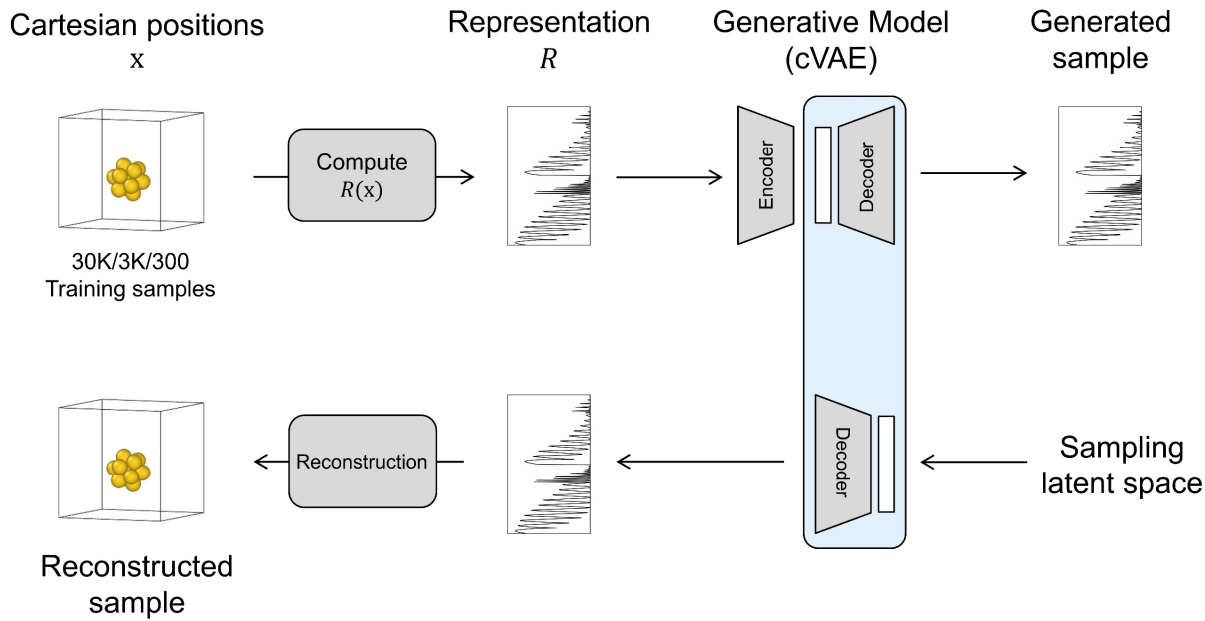

**Figure** 3: Overall scheme of the process, training, and generation steps for the end-to-end generation of new structures

The training data for the cVAE consists of a set of $Pt_{10}$ nanoclusters with various structural configurations. The structures are obtained from density functional theory (DFT) optimizations of randomly initialized geometries; both the local minima as well as intermediate structures in the optimization are used in the dataset. The target property for this dataset is formation energy in units of eV/atom, which are also obtained from the DFT calculations. Structures with a formation energy from 0 to 3.7 eV/atom were chosen. The formation energy is defined as

$$E_{form} = (E_{total} - N \cdot E_{bulk})/N \qquad (4)$$

From this overall dataset of approximately 100,000 structures, three sub-datasets are created by sampling from the total dataset using furthest point sampling based on their computed R to yield datasets containing 30,000 (Figure S1), 3000 (Figure S2), and 300 structures (Figure S3). An 80:20 training to validation ratio is used.

The DFT calculations were performed using the Vienna Ab Initio Simulation Package (VASP).[43-44] The Perdew-Burke-Ernzerhof (PBE)[45] functional was used for electron exchange and correlation energies. The projector-augmented wave method was used to describe the electron-core interaction.[43, 46] A kinetic energy cutoff of 450 eV was used. The electronic convergence criterion was set to $1\times10^{-6}$ eV. All calculations were performed without spin polarization. The Brillouin zone was sampled at the Gamma point only. A cubic cell of 17 Å was used to prevent interactions between periodic images.

## Results

We first validate the effectiveness of the proposed reconstruction method outlined in Algorithm 1 for the ACSF representation on a wide range of artificial and realistic atomic structures. Here, the aim is to obtain atomic positions which are as close to the ground truth material using only their ACSF representation. Starting with examples of reconstructing non-periodic geometric structures (Figure 4), we find the reconstruction algorithm is generally able to find solutions which are nearly identical to the ground truth. In certain cases, the reconstruction loss can be reduced to within numerical error, while in others, minor distortions can still be perceptible (structure D in Figure 4). Next, we move to reconstruction for periodic and multi-elemental structures. Periodicity is incorporated in the ACSF representation by following the minimum image convention when calculating distances and angles. For systems with more than one element, the canonical ACSF treats each atomic pair and triplet separately, and the system is then initialized with the same unit cell chemical formula as the ground truth structure. Here, we find the algorithm can obtain a nearly identical crystal structure (albeit translated in space) for the six examples in Figure 5, ranging from simple binary and ternary compounds to more complex cases containing molecular species (structure E) and layered materials (structure F). As the algorithm cannot guarantee finding the global minimum solution, ten trials with five hops each were performed for each structure, and the solution with the lowest $L(R^*, R)$ was selected. We note that while for more complex structures containing large numbers of atoms, more trials may be necessary, in practice these settings were sufficient for all the examples shown in Figure 4 and 5.

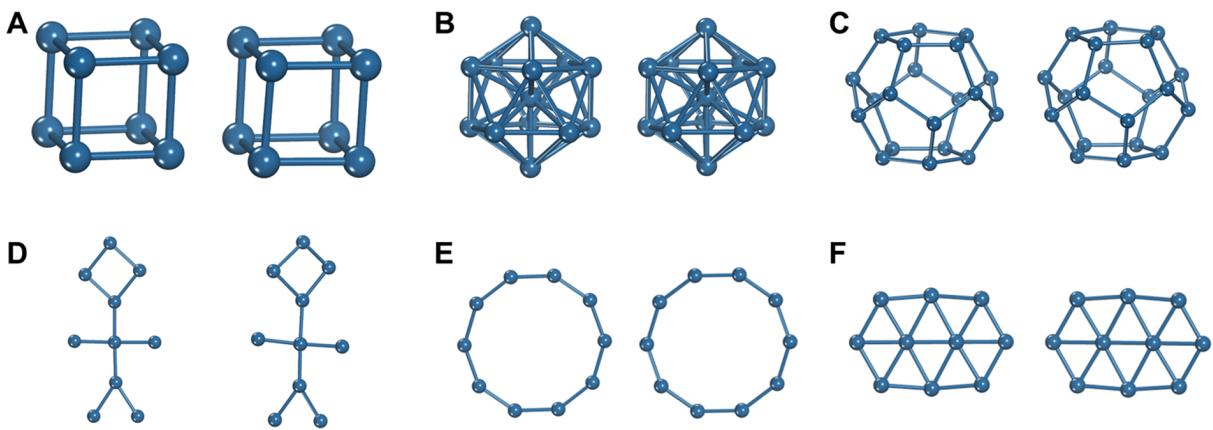

**Figure** 4: Reconstruction performance for examples of geometric structures, with the ground truth structure on the left and the reconstruction on the right. Examples shown for a (A) cube, (B) icosahedron, (C) dodecahedron, (D) humanoid, (E) decagon, and a (F) hexagonal sheet.

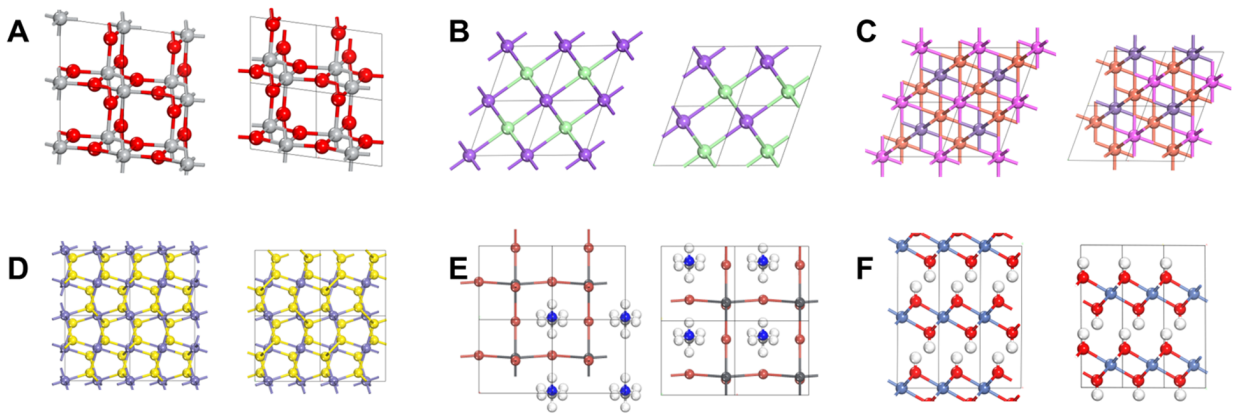

**Figure** 5: Reconstruction performance for examples of realistic crystal structures, with the ground truth structure on the left and the reconstruction on the right. Examples shown for (A) rutile $TiO_2$, (B) cubic NaCl, (C) Heusler $MnAlCu_2$, (D) pyrite $FeS_2$, (E) perovskite $CH_3NH_3PbBr_3$, and (F) layered $Ni(HO)_2$. Supercells are displayed to ease visual comparisons.

To demonstrate a fully end-to-end generation process, we formulate a task of generating structures of $Pt_{10}$ nanoparticles based on a property metric of formation energy, a measure of their thermodynamic stability. Sub-nanometer Pt nanoparticles have been studied extensively and are known to exhibit a high degree of structural fluxionality with important implications for catalysis. The formation energy of these systems is highly sensitive to even minor changes in structure and are quantum mechanical in nature, which cannot be captured by simple pair-wise interaction models resulting in complex potential energy surfaces. This therefore serves as a challenging benchmark task for structure generation due to their complex structure-property relationship and high sensitivity to errors in the generation or reconstruction process. Three training datasets of 30,000, 3000 and 300 data points are compiled from a larger dataset containing various Pt configurations and their DFT-computed formation energies.

A cVAE is then trained to serve as a generative model for new samples within the representation space R, conditional on y, the formation energy. The generated samples are then evaluated with a surrogate model trained directly on R to predict y. Table 1 shows the mean absolute error (MAE) and percent within range performance of the cVAE over 10,000 generated samples for the three training dataset sizes. Starting with the largest 30,000 dataset, the cVAE does particularly well for generation, with nearly 100% of the samples being within ±0.1 eV/atom of the targeted formation energies ranging from 1.9 to 3.7 eV/atom. To investigate the data efficiency of this approach, smaller datasets of 3000 and 300 data points are also use to train the model and compared. Remarkably, the generation performance remains strong even using only 300 training samples, and on average 80% of the generated samples falls within ±0.1 eV/atom of the target value with respect to the surrogate model. This performance suggests ACSFs to be an effective and efficient representation for generative models, which can likely be attributed to the inclusion of the translational, rotational and permutational invariances within the representation.

**Table** 1: Performance of the generated R for different training data sizes compared to a surrogate model

|  | n = 30,000 training data | | n = 3,000 training data | | n = 300 training data | |
|---|---|---|---|---|---|---|
| Target Value (eV/atom) | Mean absolute error (eV/atom) | Percent within ±0.1 eV/atom | Mean absolute error (eV/atom) | Percent within ±0.1 eV/atom | Mean absolute error (eV/atom) | Percent within ±0.1 eV/atom |
| 1.90 | 0.05 | 96.78 | 0.07 | 75.68 | 0.06 | 87.68 |
| 2.20 | 0.02 | 99.98 | 0.03 | 99.24 | 0.05 | 89.75 |
| 2.50 | 0.01 | 99.91 | 0.03 | 98.77 | 0.05 | 88.94 |
| 2.80 | 0.03 | 99.72 | 0.04 | 96.39 | 0.06 | 82.23 |
| 3.10 | 0.03 | 99.20 | 0.04 | 94.92 | 0.07 | 73.32 |
| 3.40 | 0.03 | 99.00 | 0.05 | 89.93 | 0.08 | 64.53 |
| 3.70 | 0.08 | 84.50 | 0.09 | 70.19 | 0.08 | 75.72 |

In addition to being able to generate samples which are accurate with respect to the target property, the samples must also be sufficiently unique and diverse. We plot the generated R at several selected formation energy values, taking the first two principal components from a principal component analysis (PCA). We perform this analysis for the smallest dataset with 300 training data. As shown in Figure 6, there is a large spread in the principal components of the generated samples (red squares) for y=1.9, 2.8 and 3.7 eV/atom targets, demonstrating a high degree of diversity. These samples are also unique with respect to the training data (green circles), which only sparsely covers the space of the full 100,000 dataset (black circles). This suggests the cVAE model, having learned a continuous, low-dimensional latent representation of the chemical space, can now successfully populate that space with novel samples. We note for y=1.9 many generated samples fall outside of the distribution of both the training data and the full dataset within the range of 1.8≤y≤2.0 (blue circles). These will likely result in "invalid" representations and are likely due to the low number and proportion of training samples available in that range of y, which will be discussed further.

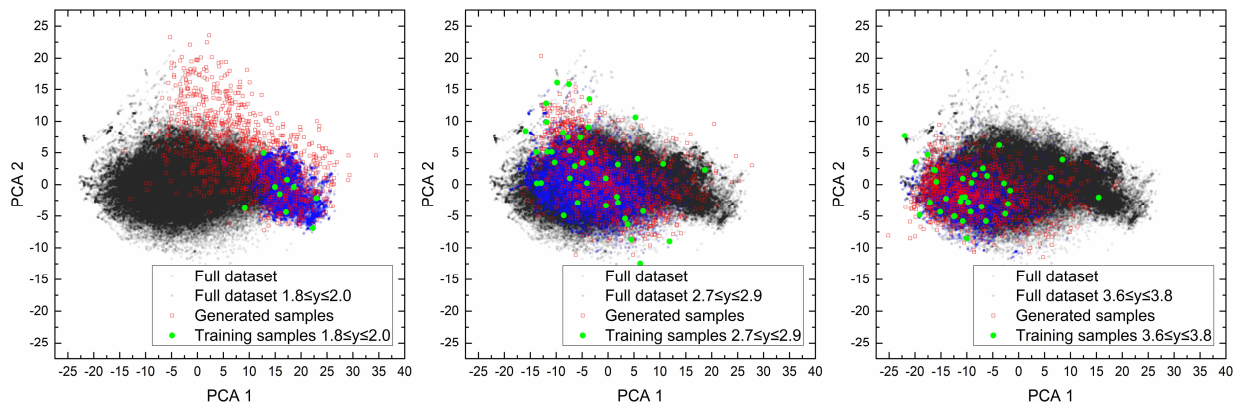

**Figure** 6: Principal component analysis of R for the dataset and generated samples at y=1.9, 2.8, 3.7

The generated representations are then reconstructed back to Cartesian positions using Algorithm 1. The reconstructed atomic structures are subsequently validated with DFT to obtain their true formation energy (y) values. No further optimization, screening or down-sampling is applied here to provide the raw performance of this generation and reconstruction process. We perform this process on 100 generated samples for y=1.9, 2.8 and 3.7 eV/atom for each of the three models, with the results shown in Table 2. We find mean absolute errors for the structures to be consistently higher once reconstructed back to their positions, as compared to Table 1. This discrepancy is likely a combination of several key factors. First, there is an inherent error or reconstruction loss when going from the representation to the structure due to finding the exact global minimum solution. Second, the MAEs obtained in Table 1 are calculated from a surrogate model trained on R, which has its own inaccuracies. Third, it is possible to generate "invalid" representations which do not encode into a valid structure. Of these factors, we hypothesize the third one to have the strongest impact, given the strong reconstruction performance (Figures 4 and 5) and the reasonable accuracy of the surrogate model (Figure S8). This was observed earlier in Figure 6 where some of the generated samples fall outside of the range of principal component values of the full dataset of valid representations (black circles). A much higher proportion of generated samples fall outside this range for y=1.9, and unsurprisingly the post-reconstruction performance is also much lower for y=1.9 than the other two target values. We also find a convergence in performance from 3,000 to 30,000 training data, where the MAE and percentage within range are roughly similar, or even slightly worse for the larger training data case. As the model, ACSF, and training hyperparameters are fixed for all three dataset sizes with an emphasis on the n=300 case, it is likely that a better performance could be obtained. However, we leave this to a future study given the current focus on data efficiency in this work.

**Table** 2: Performance of the reconstructed samples compared to DFT

| | n = 30,000 training data | | n = 3,000 training data | | n = 300 training data | |
|---|---|---|---|---|---|---|
| Target Value (eV/atom) | Mean absolute error (eV/atom) | Percent within ±0.1 eV/atom | Mean absolute error (eV/atom) | Percent within ±0.1 eV/atom | Mean absolute error (eV/atom) | Percent within ±0.1 eV/atom |
| 1.90 | 0.30 | 0.0 | 0.43 | 0.0 | 0.54 | 0.0 |
| 2.80 | 0.14 | 36.00 | 0.12 | 49.00 | 0.18 | 22.00 |
| 3.70 | 0.15 | 36.00 | 0.13 | 50.00 | 0.15 | 49.00 |

A random selection of some real and generated structures from the studies in Table 2 are visualized in Figure 7-9. For y=1.9 eV/atom, the real samples show distorted globular structures which are largely captured in the generated structures, but are also flatter with more planar motifs than the real structures. For y=2.8 eV/atom, the real samples are more diffuse and line-like (being less stable and thus having higher formation energies) with partially dissociated atoms which is also found in the generated samples. For y=3.7, the real samples are almost completely dissociated with many isolated atoms, and the generated samples follows a similar pattern. For all cases, the high diversity of the generated structures can be visually observed, with little discernible overlap in structure between the randomly chosen samples, and consistent with the PCA plots in Figure 6. A limitation in the conditional generation performance can be readily seen for y=1.9, where the generated structures lack the same degree of globularity and deviate noticeably in their computed energies from the target.

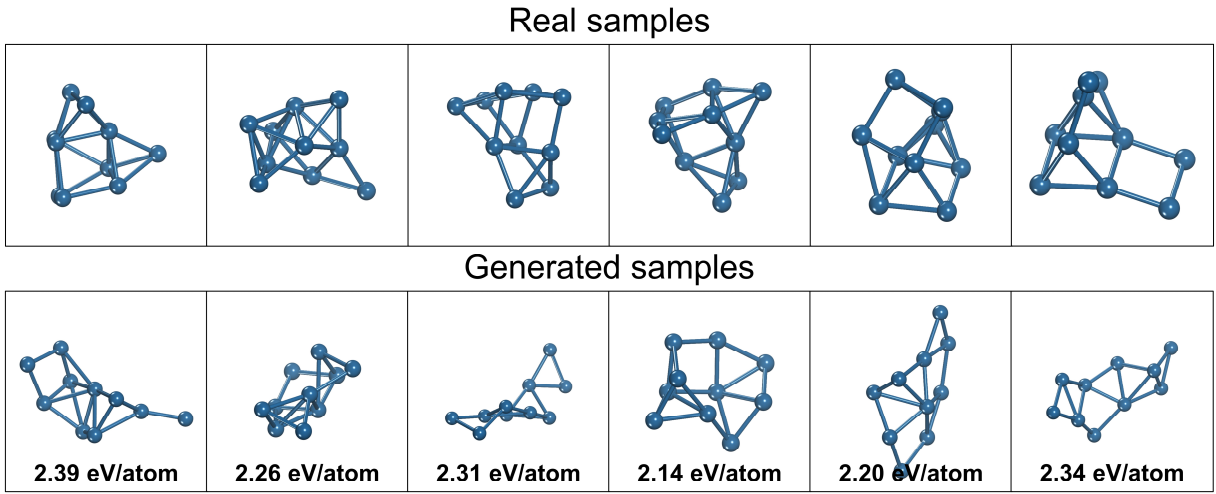

**Figure** 7: Randomly selected samples of real structures and generated structures for y=1.9

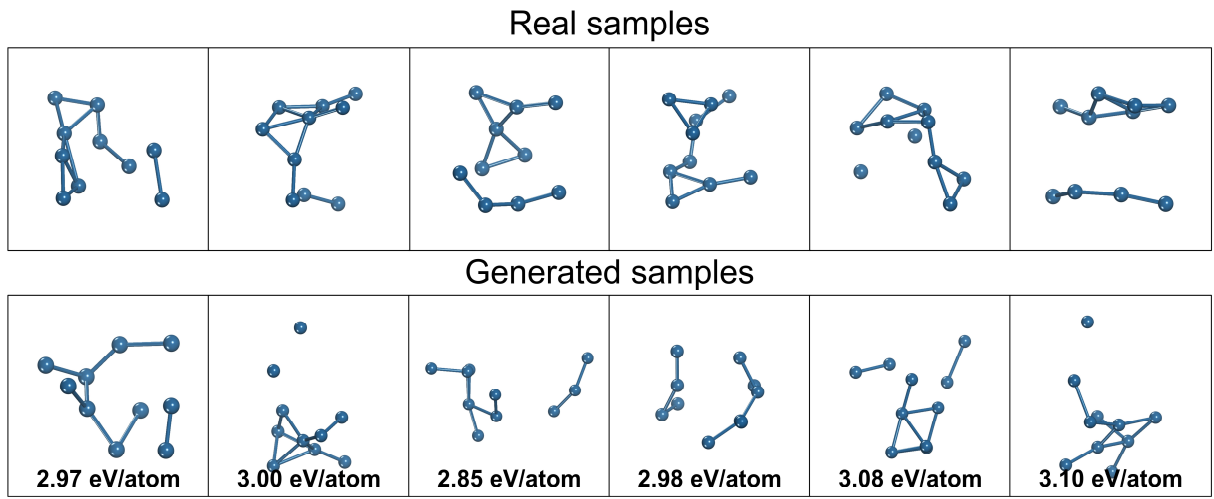

**Figure** 8: Randomly selected samples of real structures and generated structures for y=2.8

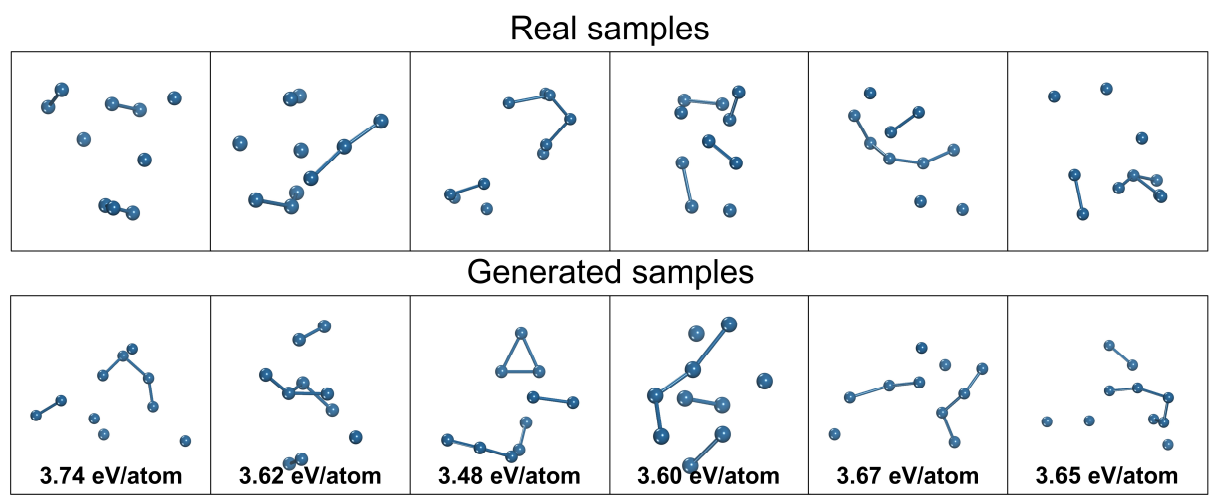

**Figure** 9: Randomly selected samples of real structures and generated structures for y=3.7

## Discussion and Conclusion

Here we provide a novel demonstration of an end-to-end generation of atomic structures following a two-step process, where materials are generated within the representation space in the first step and reconstructed to their Cartesian positions in the second step. We developed a reconstruction algorithm utilizing automatic differentiation which can be applied to any continuous and differentiable representation, where gradient-based optimization is performed find the optimal solution of positions. Using this algorithm, it is now possible, for the first time, to generate materials in the form of non-invertible representations such as ACSFs. Consequently, many other representations with the necessary structural invariances such as SOAP, ACE, FCHL, and MBTR can potentially be leveraged for generative modelling in the future. The capacity for these representations to effectively capture structural changes in an invariant manner will likely make training the generative models highly data-efficient, as demonstrated in this work by the ACSF example. In addition, the current approach is also model agnostic, such that other generative models can be easily adapted for this problem, such as generative adversarial networks (GANs), normalizing flows, energy-based models, and diffusion models which could likely provide additional improvements in performance.

Within this general approach, samples with varying atomic sizes or elemental compositions can also be generated together. A few potential approaches are envisioned to incorporate these additional dimensions in future studies. One such approach is to train the generator to provide the numbers of atoms and chemical formula in addition to the representation, and these values are then used to initialize the atoms. With this information, the reconstruction algorithm can be performed for the positions without further changes, given the current approach can already work with different sizes and elements as shown in Figure 5. Alternatively, a representation which encodes the atomic number and elemental identities can be used, and the reconstruction process can treat these quantities as variables in the optimization process itself.

Certain weaknesses may also be present in this approach due to the need to reconstruct from the representation to the Cartesian coordinates. First, the reconstruction may fail to yield a good solution with a low reconstruction loss. Although there is no upper bound to the reconstruction accuracy as present in image or voxel-based methods due to a finite grid, the optimization algorithm may still become stuck in a local minimum, particularly for systems with larger numbers of atoms. Algorithmic improvements to the global optimizer can help avoid these occurrences, however a simple solution to the problem would be to simply discard the samples with a high loss. Second, it is possible have "invalid" representations which do not encode a valid set of atomic positions, which is a well-known problem present in other representation generation methods such as SMILES to molecules. Fortunately, these cases can be readily identified by having a high reconstruction loss and can be discarded, or alternatively constraints can be placed on the generator to avoid generating invalid samples. Finally, in the current implementation, reconstruction scales poorly with system size both due to the difficulty in global optimization, as well as the increase in computational time needed for automatic differentiation in each step of the optimization. For the current example of $Pt_{10}$, and for solid state materials design problems containing under approximately 100 atoms per crystal structure, this does not present a significant problem. However, for the generation of much larger systems, further algorithmic improvements may be needed for better scalability.

We anticipate further advancements in these directions will extend this approach to generation towards the full chemical space containing both composition and structural changes in the materials. The data-efficient nature of representation-based generation can also help overcome a critical data bottleneck in solid-state materials generation, due to the sparse sampling of the chemical space (compared to molecular systems) in existing experimental or computational datasets. Structure-based generation will also be important for closing the loop in materials discovery, in concert with structural characterization techniques such as X-ray and neutron spectroscopy, diffraction, and electron microscopy.

## Acknowledgements

This work was supported by the Artificial Intelligence Initiative at the Oak Ridge National Laboratory (ORNL). Work was performed at Oak Ridge National Laboratory's Center for Nanophase Materials Sciences and used resources of the Oak Ridge Leadership Computing Facility (OLCF), which are US Department of Energy Office of Science User Facilities. VF was also supported by a Eugene P. Wigner Fellowship at Oak Ridge National Laboratory. ORNL is operated by UT-Battelle, LLC., for the U.S. Department of Energy under Contract DEAC05-00OR22725. This research used resources of the National Energy Research Scientific Computing Center, supported by the Office of Science of the U.S. Department of Energy under Contract No. DE-AC02-05CH11231.

## Competing Interests

The authors declare no competing interests.

## Additional Information

Supplementary information is available.

## Data Availability

The DFT datasets used for training are provided at https://github.com/Fung-Lab/StructRepGen.

## Code Availability

The machine learning code used in this work is available at https://github.com/Fung-Lab/StructRepGen.

# Supplementary Information

**Parameters for atom centered symmetry functions**

$G^2$ params:

    η: [0.01, 0.06, 0.1, 0.2, 0.4, 0.7, 1.0, 2.0, 3.5, 5.0]

    $D_s$: [0, 1, 1.5, 2, 2.5, 3, 3.5, 4, 5, 6, 7, 8, 9, 10]

$G^5$ params:

    *λ*: [-1, 1]

    ζ: [1, 2, 4, 16, 64]

    η: [0.06, 0.1, 0.2, 0.4, 1.0]

    $D_s$: [0]

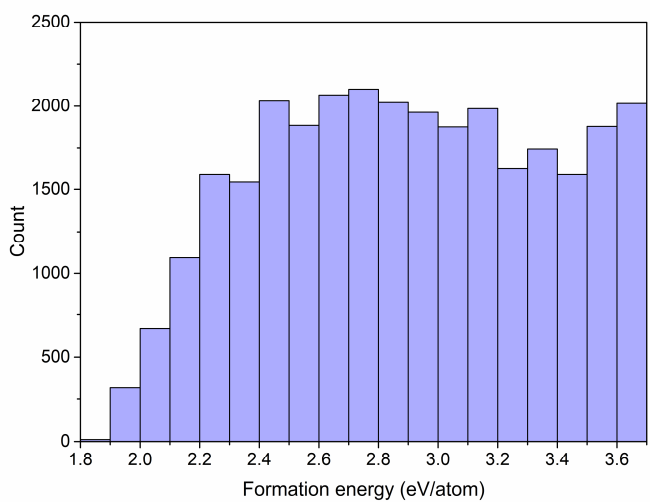

**Figure** S1: Histogram of y values for the n = 30,000 training dataset

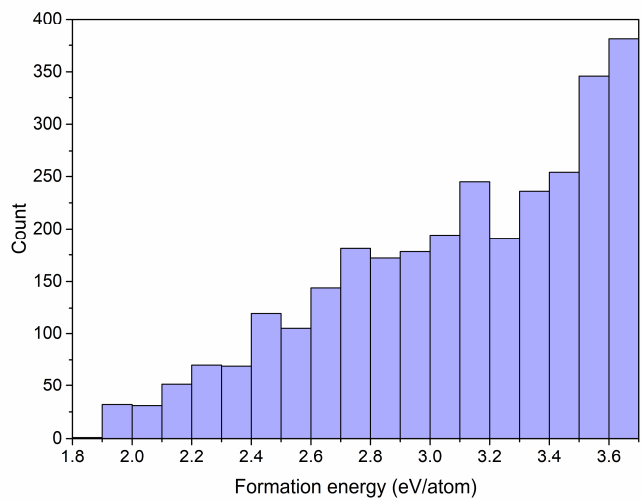

**Figure** S2: Histogram of y values for the n = 3,000 training dataset

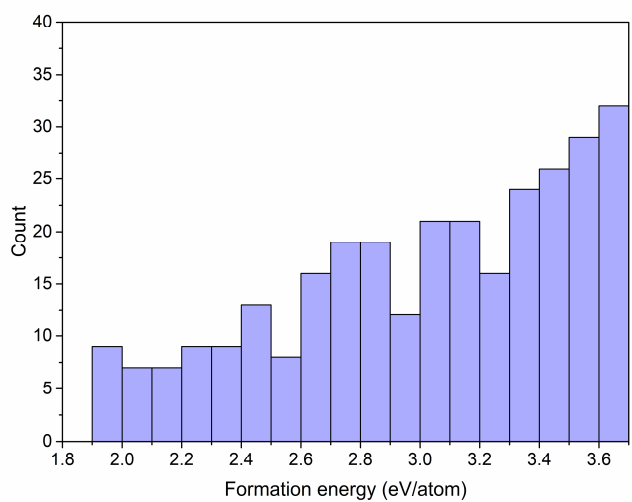

**Figure** S3: Histogram of y values for the n = 300 training dataset

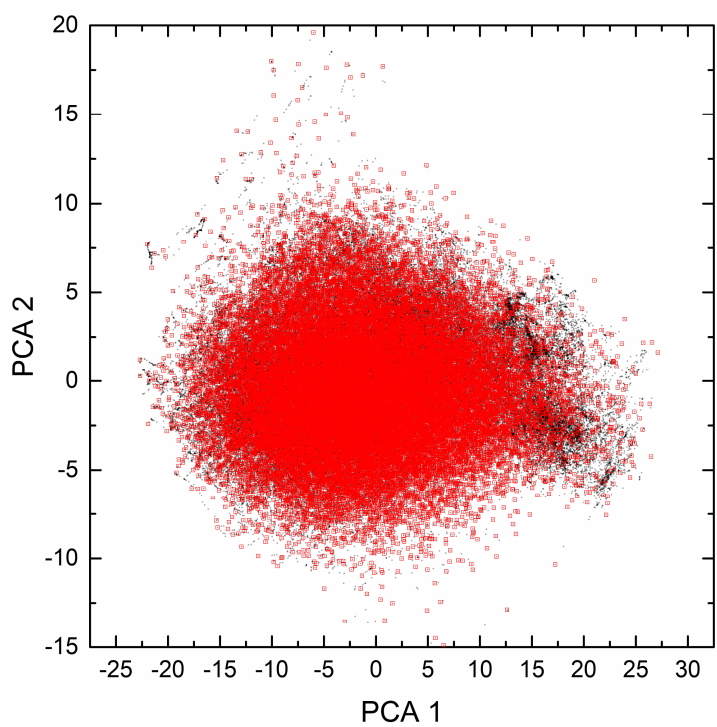

**Figure** S5: Principal component analysis of R for the full n = 100,000 dataset and the n = 30,000 training dataset

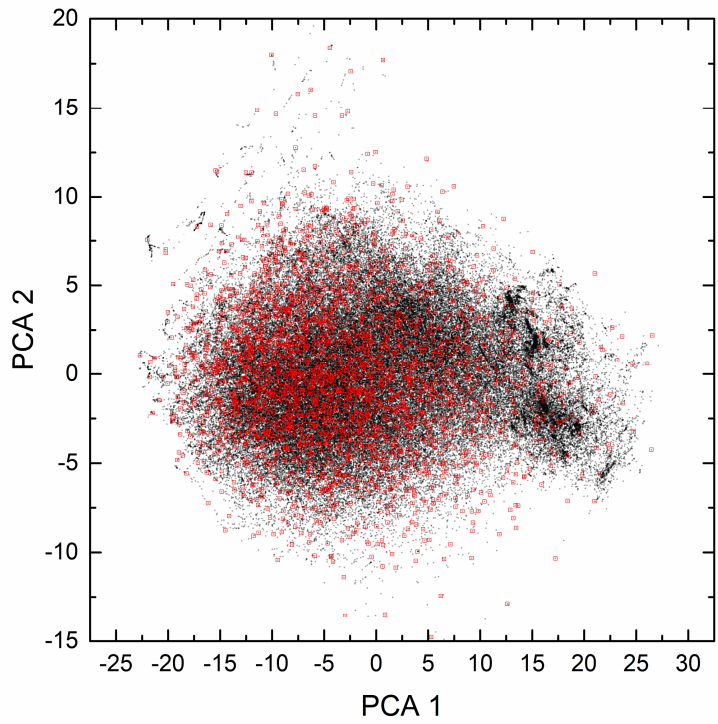

**Figure** S6: Principal component analysis of R for the full n = 100,000 dataset and the n = 3,000 training dataset

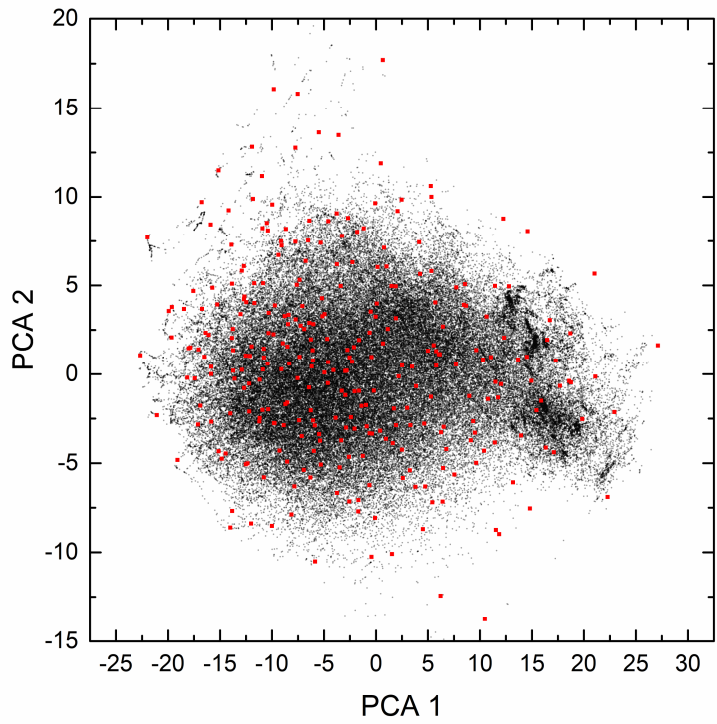

**Figure** S7: Principal component analysis of R for the full n = 100,000 dataset and the n = 300 training dataset

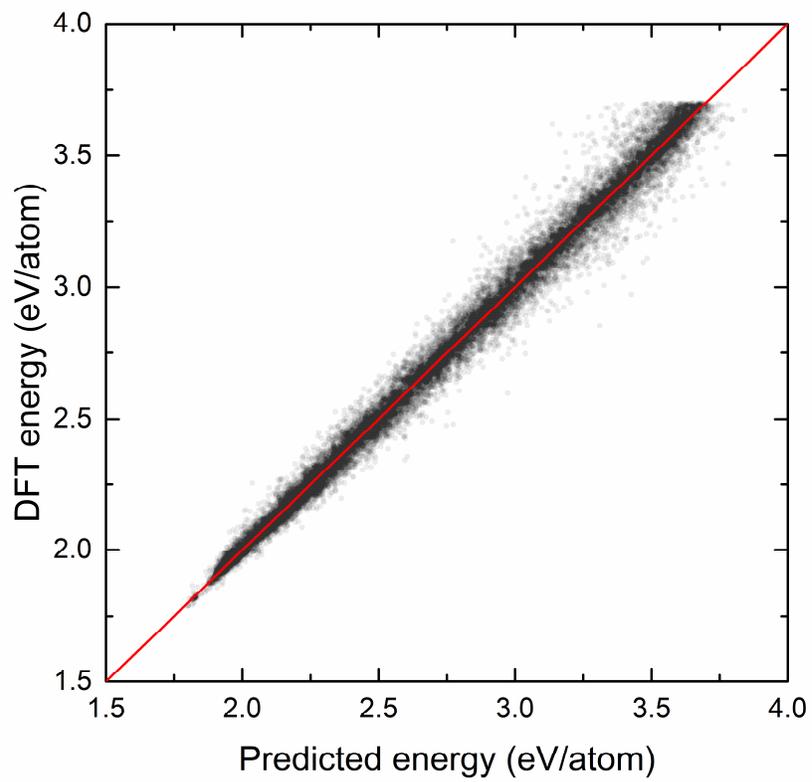

**Figure** S8: Parity plot of the predicted and actual energies of the test split for the surrogate model